# Correlated Energy-Level Alignment Effects Determine Substituent-Tuned Single-Molecule Conductance

Jeffrey A. Ivie[1,#], Nathan D. Bamberger[1,#], Keshaba N. Parida[1], Stuart Shepard[2], Dylan Dyer[1], Aldilene Saraiva-Souza[3], Roland Himmelhuber[4], Dominic V. McGrath[1,*], Manuel Smeu[2,*], and Oliver L.A. Monti[1,5,*]

[1]Department of Chemistry and Biochemistry, University of Arizona, 1306 E. University Blvd., Tucson, Arizona 85721, USA

[2]Department of Physics, Binghamton University - SUNY, 4400 Vestal Parkway East, Binghamton, NY, 13902, USA

[3]Departamento de Física, Universidade Federal do Maranhão, São Luís, MA 65080-805, Brazil

[4]College of Optical Sciences, University of Arizona, 1630 E. University Blvd., Tucson, Arizona 85721, USA

[5]Department of Physics, University of Arizona, 1118 E. Fourth Street, Tucson, Arizona 85721, USA

ABSTRACT: The rational design of single molecule electrical components requires a deep and predictive understanding of structure-function relationships. Here we explore the relationship between chemical substituents and the conductance of metal-single molecule-metal junctions, using functionalized oligophenylenevinylenes as a model system. Using a combination of mechanically controlled break-junction experiments and various levels of theory including non-equilibrium Green's functions, we demonstrate that the connection between gas-phase molecular electronic structure and in-junction molecular conductance is complicated by the involvement of multiple mutually correlated and opposing effects that contribute to energy level alignment in the junction. We propose that these opposing correlations represent powerful new "design principles," because their physical origins make them broadly applicable, and they are capable of predicting the direction *and* relative magnitude of observed conductance trends. In particular, we show that they are consistent with the observed conductance variability not just within our own experimental results, but also within disparate molecular series reported in literature and, crucially, with the trend in variability across these molecular series, which previous simple models fail to explain. The design principles introduced here can therefore aid in both screening and suggesting novel design strategies for maximizing conductance tunability in single-molecule systems.

## 1. INTRODUCTION

Modern semiconductor technology, based fundamentally on silicon, lies at a critical junction as exponentially rising costs for the development of smaller process nodes have prompted the development of alternative technologies beyond silicon.[1–3] While several promising alternative technologies have been proposed,[4–6] organic molecular electronics at the ultimate size limit of single-molecule active components offers a unique opportunity because of their structural customizability, solution-processability under benign conditions, and inherently small (~1 nm) channel lengths.[7] There has been an explosion of research into such systems in the last two decades,[8,9] resulting in single molecule devices able to reproduce the functional properties of common electronic components, including electron/hole transport selectivity through appropriate end-group selection,[10,11] rectification via molecular backbone design,[12,13] and tunability of the corresponding charge transport channels via gating.[14,15]

A significant impediment to wafer-scale fabrication of single-molecule devices, however, is a continued dearth of straight-forward, predictive, and generalizable design rules to help choose an appropriate molecular structure for achieving specific electronic properties. To help address this challenge, in this work we focus on how to modulate the fundamental electronic property of molecular conductance, $G = I / V$, by incorporating different small chemical substituents on a conserved backbone. While rarely studied to date,[16,17] variation of substituents has several advantages over other known strategies for tuning conductance,[10,18–25] including minimal impact on self-assembly, compatibility with most structures, and a near-continuous design space. In this work, the OPV3-2BT-X structure with seven different substituents is used as a model system (Figure 1a).

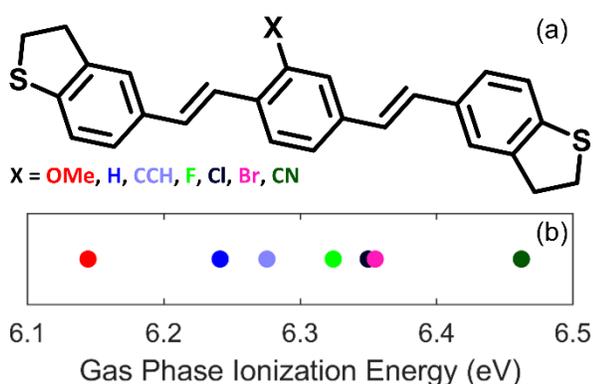

**Figure 1.** (a) Molecular structure of the OPV3-2BT-X series. The seven different substituents significantly modify the electronic structure of the molecule, as seen in particular with the gas-phase ionization energies $IE_g$ shown in (b). Note that the color convention introduced here for the different substituents will be used throughout this work.

Substituents in the OPV3-2BT-X series have a large effect on the gas-phase electronic structure of the molecule, in particular the gas-phase ionization energy ($IE_g$) values (Figure 1b). Such changes to electronic structure are expected to influence the size of the hole-injection barrier (Schottky barrier), a key determinant of interfacial charge transfer efficiency in metal-organic systems.[26] As a first approximation, one might therefore expect the large variation in $IE_g$ seen in Figure 1b to directly translate into a correspondingly large variation in the single-molecule conductance of the OPV3-2BT-X series. For example, a simple rectangular tunneling barrier model would suggest that varying the barrier height by 300 meV can easily lead to conductance variation of 50x or more. Instead, as discussed here, the conductance variation in OPV3-2BT-X and other series[16,17] is much more muted because electronic structure affects conductance via multiple interrelated mechanisms. Importantly, and as we will show, in these series the different mechanisms turn out to affect energy level alignment in anti-correlated and opposing ways, thus partially cancelling the impact of highly tunable molecular properties. A deep understanding of the physical origins of these mechanisms and their correlations with each other is therefore necessary for deducing design rules that predictively connect chemical substituents to single-molecule conductance.

To gain such an understanding, in this work we first determine the single-molecule conductance of each molecule in the OPV3-2BT-X series using both mechanically controlled break junction (MCBJ) experiments and first-principles theory combining non-equilibrium Green's functions (NEGF) and density functional theory (DFT). The limited conductance variation found by both approaches directly demonstrates the insufficiency of predicting conductance from $IE_g$ values alone. We next employ simple physical models to independently estimate the role of the two most important energy-alignment effects—namely, vacuum level shifts and image charge interactions. This reveals how the correlation of both effects with $IE_g$ explains the direction *and magnitude* of the observed conductance trend. We show that these correlations generalize to previously studied molecular systems and are even consistent with the trend across series, which $IE_g$ values alone cannot explain even at a qualitative level. This comparison across series also serves as an example of how molecular design attributes such as length can be used to modulate the strength of these correlated effects and hence control the conductance variation caused by substituents. These findings show that the two correlations described here can be used as new design rules to help identify new types of molecular scaffolds in which substituent-induced conductance variability/tailoring can be maximized.

2. RESULTS

**Experimental conductance results.** We start by presenting our experimental and theoretical conductance measurements for the OPV3-2BT-X model system, which demonstrate that substituent-induced tunability is much smaller than would be expected if one only considered the large variation in $IE_g$ values. We will then concentrate on energy level alignment effects to explain this surprising result, and discuss how correlations between those effects can be used as a new rule to connect molecular structure with observed transport properties. To experimentally characterize the single-molecule conductance of the OPV3-2BT-X series, we used an MCBJ apparatus in which the conductance of a thin gold bridge is measured while it is pulled apart to produce a so-called "breaking trace" (e.g., Figure 2a). Such breaking traces typically display a conductance plateau at ~1 $G_0 = 2e^2/h$, corresponding to a single gold atomic point contact and indicating that rupture of the gold produces atomically sharp electrodes. In the absence of molecules, the 1 $G_0$ plateau is followed by exponential conductance decay as the bridge fully breaks (e.g., blue traces in Figure 2a), but if a molecule binds inside the junction gap then a (sometimes broken) "molecular plateau" corresponding to conductance through the molecule can be observed (e.g., green traces in Figure 2a). Due to the high variation in individual breaking traces, the process is repeated thousands of times and the results are summarized in two-dimensional (2D) or one-dimensional (1D) histograms (Figure 2b,c,e). For each member of the OPV3-2BT-X series, deposition of the molecule led to the appearance of a plateau-like feature (e.g., Figure 2c) not seen in the empty-junction traces (e.g. Figure 2b), and corresponding to a broad conductance distribution centered at around $10^{-4}$ $G_0$ (e.g., Figure 2e).

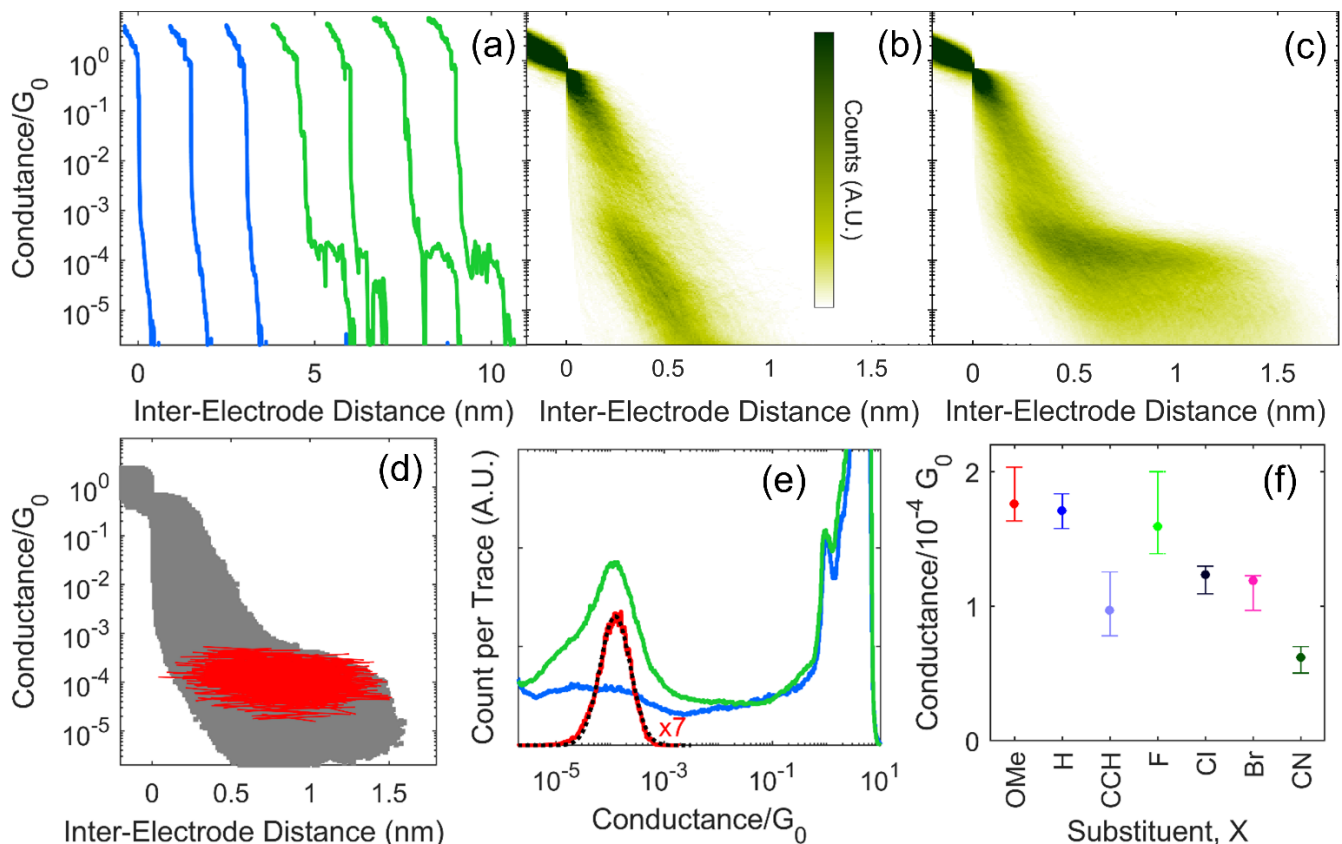

**Figure 2.** (a-e) Experimental results for one OPV3-2BT-Br dataset. (a) Selected breaking traces from before (blue) and after (green) molecular deposition, illustrating tunneling decay and (sometimes broken) molecular plateaus, respectively. (b) 2D histogram of 3468 consecutive breaking traces collected before molecular deposition, showing an exponential decay feature. (c) 2D histogram of 8603 consecutive breaking traces collected after depositing OPV3-2BT-Br, showing an extended molecular plateau feature at ~$10^{-4}$ $G_0$. (d) Segments assigned to the "main plateau cluster" for the dataset in (c), with the raw dataset distribution included in gray. (e) 1D conductance histograms for the tunneling traces from (b) (blue), the molecular traces from (c) (green), and the segments from the main plateau cluster in (d) (red) along with a Gaussian fit (dotted black). The segment distribution, normalized to the number of traces in the *entire* dataset, has been scaled up by a factor of 7 for clarity. (f) Peak conductance values for the OPV3-2BT-X molecules studied in this work, with the error bars representing uncertainty in cluster bounds and variation between multiple datasets for each molecule.

The breadth of these conductance distributions is large compared to the variation in the "background" signature of broken or incomplete plateaus (see, e.g., green traces in Figure 2a), making it challenging to differentiate conductance across the series. We therefore used a segment clustering algorithm described previously[27] to robustly extract and fit the main plateau feature from each OPV3-2BT-X dataset (e.g., Figure 2d,e). This process was repeated for multiple independent datasets for each molecular species, yielding reproducible and distinguishable peak conductance values for each molecule (Figure 2f), despite the rather broad distributions in the raw data. These measurements reveal that the conductance variation due to substituents is surprisingly small relative to the variation in gas-phase electronic structure (Figure 1b).

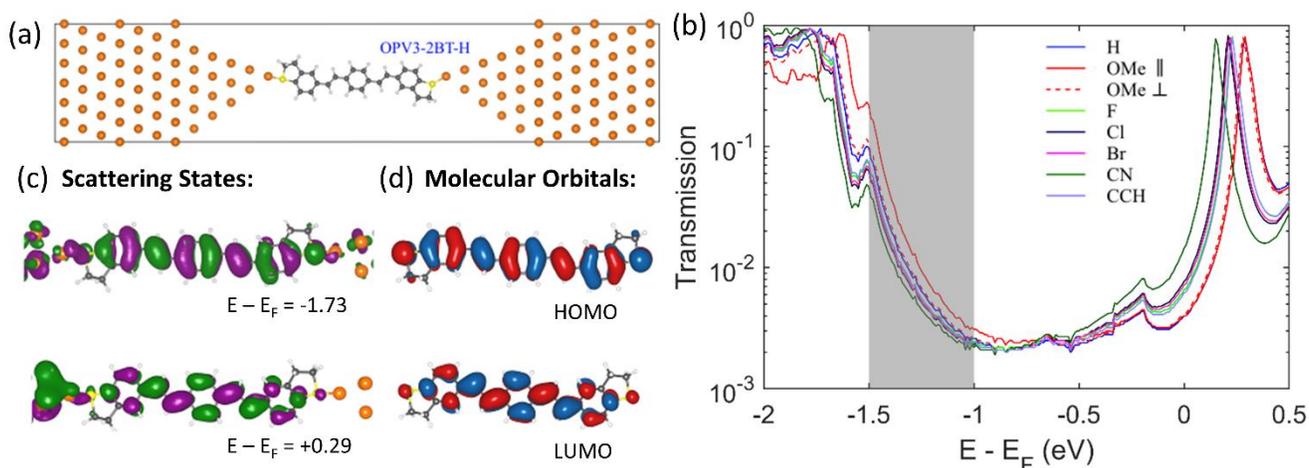

**Figure 3.** (a) Geometry of optimized electrode-molecule-electrode system with OPV3-2BT-H. (b) Comparison of calculated transmission functions for each OPV3-2BT-X molecule. Shaded gray region indicates the range in which the true $E_F$ likely falls. (c) Scattering states of the OPV3-2BT-H junction at the two transmission peaks seen in (b). (d) HOMO and LUMO orbitals of the isolated OPV3-2BT-H molecule. Comparison of (c) and (d) indicates that the two transmission peaks correspond to HOMO- and LUMO-mediated transport, respectively.

**Comparison with theoretical results.** To gain further insight and help validate our experimental findings, we turned to quantum transport calculations to determine the conductance of the OPV3-2BT-X series from first principles. These calculations combine the NEGF technique and DFT, as implemented in the Nanodcal code.[28,29] The electron many-body effects are treated in the exchange-correlation functional form of Perdew, Burke, and Ernzerhof (PBE).[30] In our simulations, each molecule is placed between gold electrodes tapering to single atom apexes (Figure 3a) at an optimal gap size of 2.3 nm (see SI section S2). While different conformations of the OPV3 backbone have minimal impact on calculated electronic structure, the orientation of the methoxy substituent in OPV3-2BT-OMe meaningfully influences the theoretical results (SI section S3). Two possible stable orientations are thus considered throughout this work: the methoxy substituent parallel to the plane of the ring (always shown in solid red), and the methoxy substituent perpendicular to the backbone (always shown in dotted and/or unfilled red).

The transmission functions for all molecules (Figure 3b) are broadly similar, with shared peaks near -1.75 eV and +0.25 eV. Comparison of scattering states at those energies (Figure 3c) with the orbitals of the isolated OPV3-2BT-H molecule (Figure 3d) reveals that those peaks correspond to HOMO- and LUMO-like transport levels, respectively. NEGF-DFT places the Fermi energy ($E_F$) on the edge of the LUMO peak, but DFT is well-known to underestimate the HOMO-LUMO gap and often misplaces $E_F$ within that gap.[31] Instead, both orbital symmetry[32] and current-voltage measurements[33] suggest that the BT linker group leads to HOMO-mediated transport in the OPV3-2BT-X series. This assignment is also supported by the fact that the ordering of transmission functions along the edge of the LUMO-peak (Figure 3b) is in stark disagreement with our experimental results. For this work we therefore place the true $E_F$ along the trailing edge of the HOMO-peak. Throughout this region (shaded in Figure 3b), the ordering of molecular conductances remains consistent, allowing for a direct comparison to experimental results. Although quantitative agreement between calculations and experiment is not expected at this level of theory because calculated conductances tend to overestimate measured values by an order of magnitude,[33] we expect NEGF-DFT to capture the correct relative conductance trends.[34,35]

Experimental and theoretical results are compared in Figure 4, with both sets of values normalized by the conductance of OPV3-2BT-H to remove effects that might arise from DFT systematically under-estimating band-gaps.[36,37] Broad agreement between experiment and theory is observed, both in terms of the qualitative ordering of the molecular conductances *and* in the relative magnitude of conductances. Both theoretical and experimental results thus agree that small substituents can be used to tune molecular conductance in the OPV3-2BT-X series, as is expected due to the variation in gas-phase electronic structure properties including $IE_g$. The major finding, however, is that the magnitude of this tunability is surprisingly small given the range of $IE_g$ (i.e., a factor of 2 to 3 change in conductance vs. a range of $IE_g$ in excess of 300 meV). In the remainder of the paper we show how this strikingly small sensitivity of conductance to molecular electronic structure stems from a peculiar anti-correlation of several different effects, and show that this understanding of energy level alignment in the junction provides a unified picture of conductance trends in diverse molecules beyond OPV3-2BT-X. This insight defines new possibilities for tuning energy level alignment and hence conductance in molecular junctions.

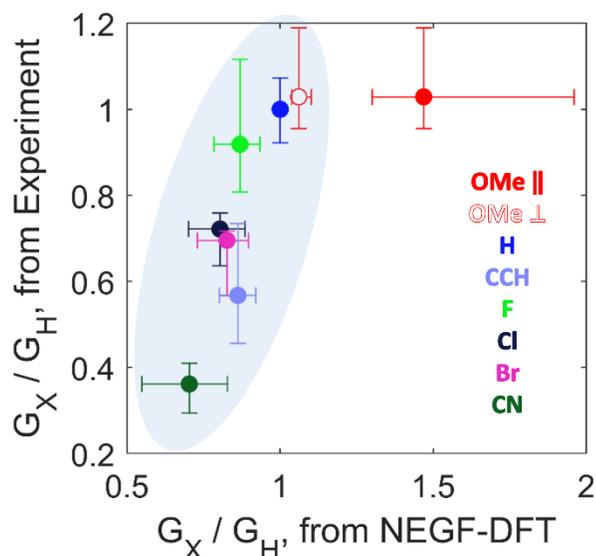

**Figure 4.** Comparison of experimental and theoretical conductances for the OPV3-2BT-X series ($G_X$), relative to the conductance of OPV3-2BT-H ($G_H$). Experimental error bars represent uncertainty arising from independent datasets collected for each molecule and from uncertainty in cluster bounds, while theoretical error bars represent uncertainty about where in the shaded region in Figure 3b the true $E_F$ lies (see details in methods section). The shaded oval highlights the close linear agreement between experiment and theory for all points except –OMe in the parallel orientation; however, even including –OMe-parallel, there is broad qualitative agreement in the ordering of conductances.

3. DISCUSSION

**Conceptual model of level alignment effects.** In order to understand how multiple anti-correlated effects explain the limited variability of observed conductance across the OPV3-2BT-X series, we first propose a simple conceptual model that allows us to consider the contributions of each major effect separately. The molecules in the OPV3-2BT-X series share the same conjugation, length, and linker group, and calculations suggest no meaningful variation in the twist angles between oligomer units (SI section S4). We therefore focus our attention on energy level alignment, as this is the primary attribute in which these molecules are expected to differ from each other. In particular, since the OPV3-2BT-X series is expected to be HOMO-conducting, variation in the hole-injection barrier (HIB) should determine the conductance trend across these molecules.

Figure 5 illustrates the three major physical mechanisms that control variation in HIB values in our conceptual model. The starting point in the limit of no molecule-electrode interactions is the molecular gas-phase ionization energy $IE_g$ referenced to the vacuum level of the empty junction ($V_{EJ}$) (Figure 5, panel 1; vacuum level alignment or Schottky-Mott limit). As is well known from organic/metal interfaces, coupling of a linker group to the electrode changes the local vacuum level due to the build-up of a surface dipole.[38] This typically shifts $V_{EJ}$ down by $\Delta V$, moving the transport level farther from $E_F$ (Figure 5, panel 2). Finally, interactions between the charge distribution of the positively charged ionized molecule and its image charge distribution in the electrodes renormalizes the transport levels;[39] this renormalization energy (RNE) reduces the ionization energy and thus decreases the HIB (Figure 5, panel 3). While this overall picture of energy level alignment does not include effects such as surface reconstruction, such higher-order contributions are expected to play a minor role and to be mostly conserved across the series as long as the linker group is maintained. The three effects listed in Figure 5 thus represent the primary determinants of relative molecular conductance in our system.

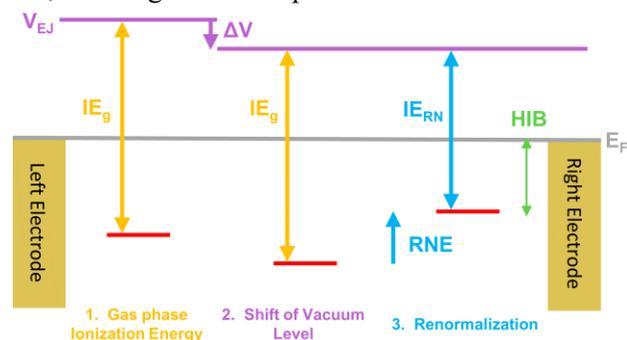

**Figure 5.** Energy level diagram illustrating the major effects that control conductance variation across the OPV3-2BT-X series by determining the size of the hole injection barrier (HIB, green). The starting point is the alignment between the gas phase ionization energy ($IE_g$, yellow and referenced to the vacuum level of the empty junction, $V_{EJ}$) and the Fermi energy of the electrodes ($E_F$; panel 1). Interactions between the molecule and the electrodes cause $V_{EJ}$ to decrease by $\Delta V$ (purple), moving the transport level (red) farther from $E_F$ (panel 2). Interaction between the molecular charge distribution and image charges in the electrodes then causes the ionization energy to shrink by RNE (blue), reducing the HIB (panel 3). In the OPV3-2BT-X series, these three effects are correlated such that any changes to $IE_g$ are partially canceled out by the changes that simultaneously occur to $\Delta V$ and RNE (see Figure 6a,b).

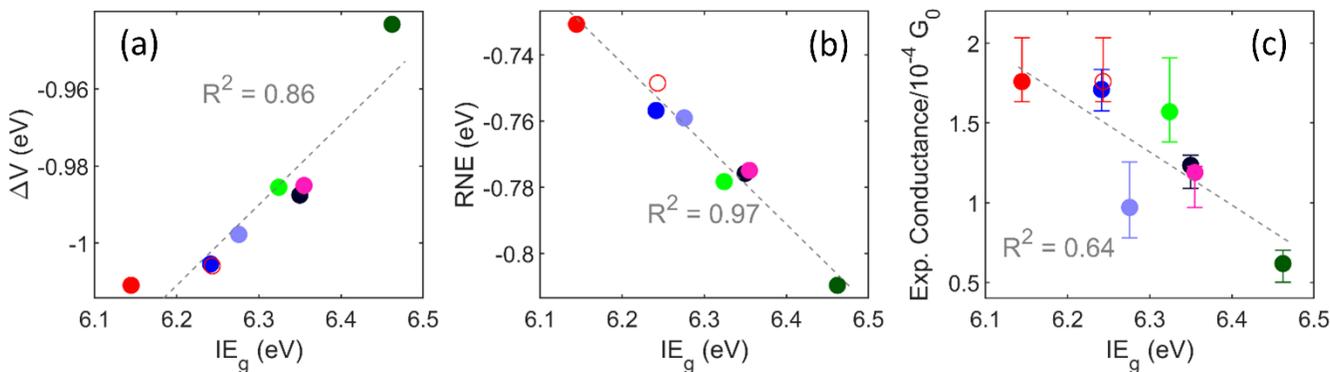

**Figure 6.** Correlations between different level-alignment and conductance values for the OPV3-2BT-X series. (a) Correlation between calculated values of $\Delta V$ and of $IE_g$. (b) Correlation between calculated values of RNE and of $IE_g$. (c) Correlation between experimentally measured conductances and calculated $IE_g$ values. The fact that $\Delta V$ and RNE are each correlated with $IE_g$ explains both why $IE_g$ alone predicts the qualitative trend in conductances quite well, and why the variation in conductance is so minimal.

To investigate the OPV3-2BT-X series in accordance with the conceptual model in Figure 5, we employ a local electrostatic potential model to estimate $\Delta V$ values and a simple image charge model to estimate RNE values (see methods section for details of both models). These two models of specific effects complement our NEGF-DFT model: whereas the latter provides a single integrated quantitative picture, the former provide conceptual insight into the physical origins of observed trends. As shown in Figure 6a,b, these independent calculations reveal a tight linear correlation between $IE_g$ and both $\Delta V$ and RNE. These correlations consequently explain why the conductance trend is broadly predicted by $IE_g$ values alone (Figure 6c), despite there being three major components that determine HIB values (Figure 5). Moreover, the directions of the correlations in Figure 6a,b are such that increasing $IE_g$ is associated with a less-negative $\Delta V$ and a more-negative RNE. This leads to our main conclusion: variation in both $\Delta V$ and RNE opposes any variation in $IE_g$, leading to a much smaller spread of both HIB values and conductances than might otherwise be expected. In particular, our calculations suggest that of the ~320 meV spread in $IE_g$ values across the OPV3-2BT-X series, variation in $\Delta V$ counteracts ~70 meV of this spread and variation in RNE counteracts ~80 meV. As these numbers are the result of two different simple models, they should only be taken as estimates. However, the fact that all three ranges are of the same order of magnitude provides strong evidence that the correlations delineated in Figure 6a,b play a significant role in reducing the range of HIB values in the OPV3-2BT-X series, and are thus the primary reason for the direct correlation of conductance with $IE_g$ *and* the limited conductance tunability observed in Figure 6c.

In the following sections, we probe the $\Delta V$ and RNE models independently to uncover the causes of the correlations in Figure 6a,b. We show that these correlations are not happenstance, but arise instead due to inherently linked physical phenomena, and therefore serve as new design principles that are expected to apply to a broad range of molecular structures. To demonstrate this wide applicability, we show in the final section how the RNE correlation in particular yields a unified description of previously published results.

**Correlation of vacuum level shifts.** To explain the correlation between $IE_g$ and $\Delta V$, we decompose the vacuum level shift into two components (Figure 7a). First is the push-back effect,[40,41] $\Delta V_{PB}$ (Figure 7a, panel 1). We assume this contribution to be constant across the series because push-back should mainly be caused by the shared BT linker group. Next, coupling to the electrodes broadens the molecular energy level, resulting in charge-transfer from the tailing molecular states extending above $E_F$ to the electrodes (Figure 7a, panel 2). The resulting surface dipole produces the second contribution, $\Delta V_{CT}$ (Figure 7a, panel 3).[42] Electrostatic potential calculations (see methods section and SI section S5 for details) in the junction show that $\Delta V$ exhibits a tight linear correlation with the amount of charge transferred off the molecule (Figure 7b), which we attribute to variation in $\Delta V_{CT}$.

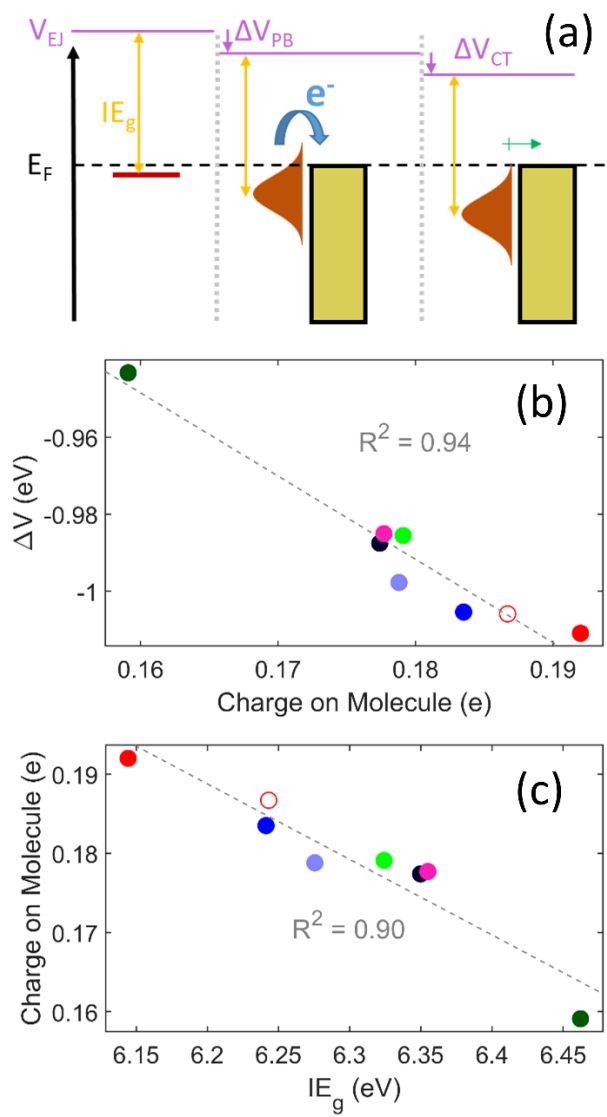

**Figure 7.** (a) Conceptual model of physical origins of the local vacuum level shift $\Delta V$. A portion of this shift, $\Delta V_{PB}$, is due to the push-back effect. Charge transfer between the broadened molecular energy level and the electrodes then causes a further shift $\Delta V_{CT}$. (b) A tight correlation is found between $\Delta V$ and the charge on the molecule; this is attributed to the contribution of $\Delta V_{CT}$, which is proportional to the amount of charge transfer. (c) The amount of charge transferred off the molecule is in turn tightly correlated with $IE_g$, thus explaining the link between $IE_g$ and $\Delta V$ seen in Figure 6a.

The molecular charge transfer is in turn tightly correlated with $IE_g$, because $IE_g$ determines what fraction of the broadened molecular energy level extends above $E_F$ upon junction formation (Figure 7c). This is the same relationship that can lead to Fermi-level pinning in more extreme cases where the molecular energy level is brought especially close to $E_F$.[43,44] The correlation between $\Delta V$ and $IE_g$ seen in Figure 6a is thus mediated by the fundamental physical process of interfacial charge transfer, and so should apply to most HOMO-conducting molecular systems. Even in the case of LUMO-conducting molecules, however, an analogous opposing correlation is expected due to charge transfer from the electrodes into the broadened LUMO level.

**Correlation of image charge renormalization.** To understand the correlation between RNE and $IE_g$, we first conceptually simplify our view of the image charge effects to focus on the most impactful interactions. The full calculation of RNE in our model (see methods section for details) involves summing thousands of individual pair-wise interactions between the original partial charges on the molecule and hundreds of sets of image charges in both the neutral and cationic molecule (Figure 8a). As explained in SI section S6, we find that 1) the dominant interactions are those in the cationic molecule involving the first order image charges of the hydrogens; and 2) like atoms can be grouped together. This simplified picture is illustrated in Figure 8b for the case of OPV3-2BT-H: a modest repulsive potential exists between the negatively charged carbon atoms and the negatively charged hydrogen images, but this is more than counteracted by a large attractive potential between the positively charged hydrogen and sulfur atoms and the negatively charged hydrogen images, explaining why the RNE for OPV3-2BT-H has an overall negative value of -0.76 eV.

This reduced image charge picture can now be used to understand trends in RNE *across* the OPV3-2BT-X series. For example, for a change from –H to a more electronegative substituent such as –CN, the substituent itself becomes more negatively charged at the expense of the atoms of the OPV3 backbone (Figure 9a). As shown in Figure 9b,c, while this extra negative charge on the substituent increases repulsion with the hydrogen images, at the same time the more positive backbone causes increased attraction with the hydrogen images. Crucially, because it is closer to the electrodes on average, the backbone response dominates. This means that making the substituent more electron-withdrawing directly causes RNE to become more negative, and an analogous argument shows that more electron-donating substituents cause an increase in RNE (SI section S10). Since the substituents are directly connected to the molecular π-system, substituent donating or withdrawing character also directly controls $IE_g$,[17] leading to the tight RNE vs. $IE_g$ correlation shown in Figure 6b. This physical link can therefore be expected to be present in a wide variety of molecular systems, and an analogous anti-correlation also arises in the case of LUMO-mediated transport (SI section S11). In the next section, we explicitly show how this explanation unifies single molecule conductance observations for other seemingly unrelated systems.

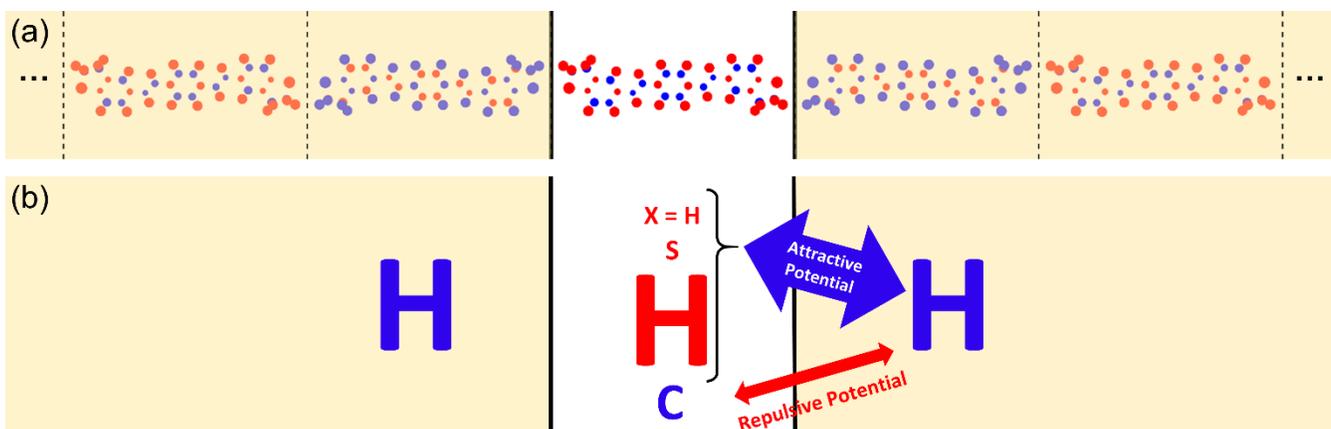

**Figure 8.** (a) Representation of the atomic partial charges on the OPV3-2BT-H cation, with red (blue) circles indicating positive (negative) charges and circle area proportional to charge magnitude. The planar electrodes are also represented (gold) along with multiple sets of charge images induced within each electrode. The full calculation of RNE for OPV3-2BT-H involves contributions from all of these image charges, as well as from those of the neutral molecule (not shown). However, as justified in SI section S6, at a conceptual level we can simplify this picture to the cartoon in (b), in which atoms are grouped by type, neutral molecule interactions are ignored, and we focus solely on cation interactions involving the (negatively charged) first order hydrogen images. This demonstrates that the negative sign of RNE for OPV3-2BT-H arises because attractive interactions are dominant.

**Extension to other molecular systems.** Based on the physical connections discussed above, both of these design principles—the correlation of both $\Delta V$ and RNE with $IE_g$—should extend to a broad variety of molecular systems beyond OPV3-2BT-X, where they may play a similarly important role in determining the range of conductance variation. To investigate this generalizability, we focus here on the RNE effect since it is more readily computationally accessible and more distinct from previously reported relationships in multi-molecule junctions.[45]

We concentrate on the two most prominent additional cases from the literature in which single-molecule conductance was measured across a series of molecules with identical linker groups and backbones but varying substituents: the benzene diamine (BDA) series studied by Venkataraman et al.[17] (Figure 10a), and the penta-phenyl cyclophane dithiol series (P5-cyclophane-2SH-X) studied by Lo et al.[16] (Figure 10b). These series span a wide swath of design space, as they differ in length, backbone structure, and linker group compared to both OPV3-2BT-X and each other. However, after applying our model to both series, we find the same tight anti-correlation between $IE_g$ and RNE as found previously in the OPV3-2BT-X series (Figure 10). We therefore predict that—just as for OPV3-2BT-X—the conductances in these other series are qualitatively well-explained by $IE_g$ values alone, but that the magnitude of conductance variation is small relative to the variation in $IE_g$. As shown in SI section S12, this prediction is consistent with the conductance measurements reported for the BDA and P5-cyclophane-2SH-X series. This demonstrates that the opposition between $IE_g$ and RNE that we discovered in the OPV3-2BT-X series is quite general, predictive, and provides a uniform theoretical underpinning for understanding the direction and magnitude of conductance trends *within* diverse molecular series.

We now consider the trend in conductance variability *between* the three series (BDA, OPV3-2BT-X, and P5-cyclophane-2SH-X). We note that focusing on $IE_g$ values alone, which works well to explain conductance ordering within series,[16,17] would predict exactly the wrong trend in relative conductance between the three series: the largest $IE_g$ variation is in fact associated with the *least* relative conductance variation (Table 1; we focus on the substituents –Cl and –OMe since they are included in all three series). In contrast, considering RNE values as well is fully consistent with the observed conductances, because the series with more RNE variation—which is predicted to reduce conductance variability by opposing shifts in $IE_g$—do indeed display less variation in relative conductance (Table 1). This trend in RNE variation appears to be primarily a consequence of backbone length (see Table 1 and SI section S13), demonstrating how the $IE_g$/RNE correlation serves as a design principle that directly connects molecular structure to transport properties. Due to model limitations, the $IE_g$ and RNE ranges in Table 1 cannot be directly compared to determine which trend across series dominates, and of course these three molecular series differ in other ways that may also influence conductance variability. However, the fact that the $IE_g$/RNE correlation provides a unified view of conductance across diverse molecular designs that is consistent with the conductance trend *between* series illustrates how this design principle offers powerful insight into single-molecule conductance trends.

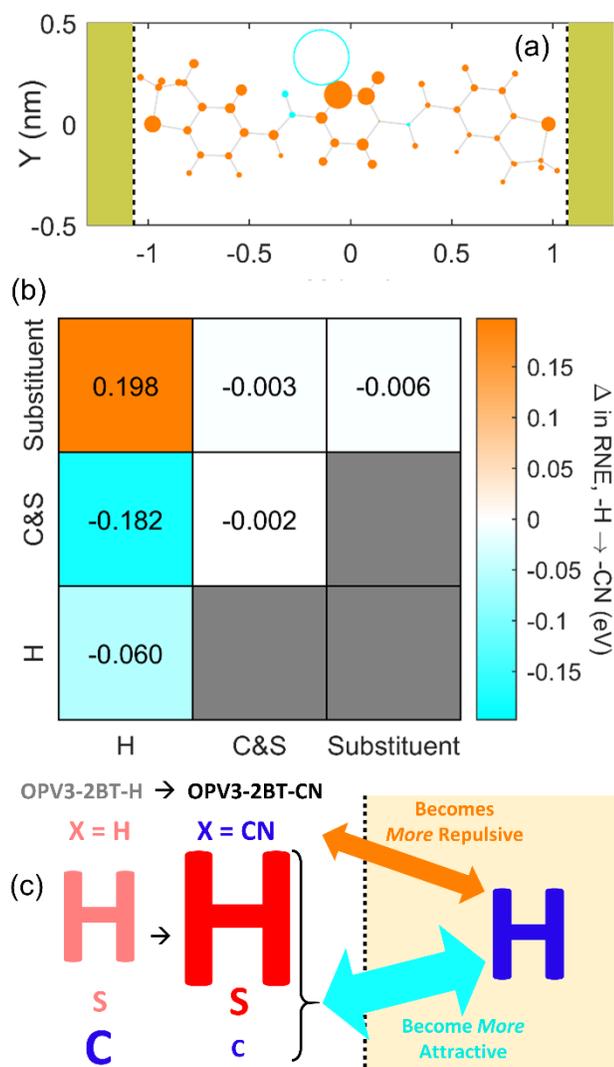

**Figure 9.** Changes in image charge interactions when the –H substituent is replaced with –CN. (a) Change in partial charge on each conserved atom, with orange (cyan) circles indicating charges that became more positive (negative), and the area of each circle proportional to the magnitude of change. The net charge change on the substituent is represented by the open circle. (b) Changes in cation contributions to RNE, with atoms grouped into three categories: conserved hydrogens, conserved carbons/sulfurs, and atoms belonging to the substituent. Energies have been summed across the diagonal due to the symmetry of the image charge interactions (SI section S7). These energy changes can be rationalized using the simplified cartoon model in (c): the substituent net charge flips from positive to negative, leading to more repulsive interactions with the hydrogen images. However, this is outweighed by the increase in attractive potential due to the backbone atoms becoming more positively (or less negatively) charged.

## 4. CONCLUSIONS

In this work, we have investigated the use of small chemical substituents as a means to tune single-molecule conductance, using the OPV3-2BT-X series as a model system. Our results confirm that such substituent-induced tuning is indeed possible. While $IE_g$ values alone superficially appear to explain the *trend* in conductance variation, we demonstrated that they are insufficient to explain the variation in *magnitude*. Instead, the superficial $IE_g$ trend conceals a more intricate relationship in which variation in vacuum level shift ($\Delta V$) and image charge renormalization energy (RNE) also significantly affect conductance, but these latter two energies turn out to be highly correlated with, and in opposition to, variation in $IE_g$. Using simple physical models, we showed how these correlations arise due to electrostatics in the junction providing inherent physical links, and how the degree of opposition can be controlled by molecular structure. We therefore propose that these correlations serve as new "design principles" that are crucial for explaining, predicting, and controlling conductance variation across molecular series with varying substituents, in the case of either HOMO- or LUMO-mediated transport. This generalization was demonstrated by extending our calculations to previously measured molecular series, and showing that the predictions of the RNE design principle in particular are consistent with the trends both within and between those series.

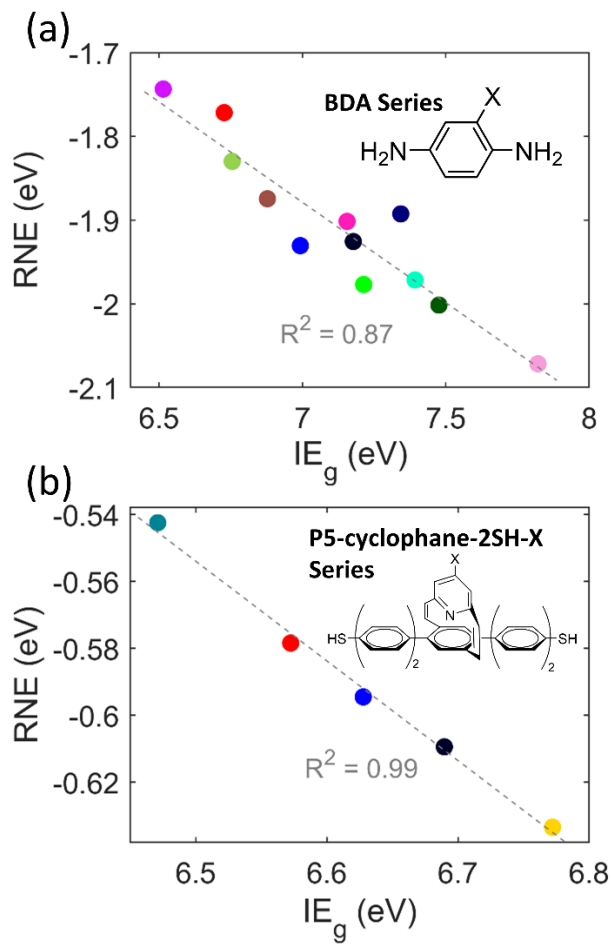

**Figure 10.** Anti-correlation between calculated RNE and $IE_g$ values for the BDA series studied by Venkataraman et al.[17] (a) and the P5-cyclophane-2SH-X series studied by Lo et al.[16] (b), demonstrating that the relationship from **Error! Reference source not found.**c generalizes beyond OPV3-2BT-X. See SI section S12 for list of substituents considered for each series (note that a few molecules in the BDA series have multiple substituents).

In the molecular series studied here, the $\Delta V$ and RNE correlations serve to oppose changes in $IE_g$ and hence limit conductance variation between molecules. Looking to the future, the question naturally arises of whether it is possible to break or circumvent these correlations, as this would be desirable from the perspective of maximizing substituent-induced conductance tunability. The prospects for avoiding opposition from $\Delta V$ appear limited, especially if high absolute conductances are also desired. Eliminating the extension of the broadened molecular energy level above $E_F$ would require either lowering molecule-electrode coupling or significantly increasing $IE_g$, both of which would drastically diminish total conductance. However, it may still be possible to reduce $\Delta V$ opposition somewhat while maintaining reasonable absolute conductance, and local electrostatic potential calculations of the type performed here can be used as a screening tool to search for structures that achieve this.

We propose that overcoming the RNE correlation is a more promising strategy, because image charge effects are also sensitive to substituent positioning and connectivity. We note for example that while the three series discussed in this work are quite different in many respects, they also have several attributes in common: they are all based on conjugated backbones, they all use relatively small substituents, and those substituents are all attached directly to the center of the backbone. More exotic molecular scaffolds which do not share one or more of these properties may therefore allow the $IE_g$/RNE anti-correlation to be effectively eliminated, or even reversed. The simple image charge model described here, together with our conceptual model of the dominant image charge interactions, can each help in both searching and screening for such novel structures. This work thus demonstrates how a deep understanding of the physical origin of the connections between molecular structure and electronic properties is essential for the intentional design of new functional single-molecule devices in the future.

**Table 1. Comparable[a] Conductance Ratios and Energy Ranges for the Three Molecular Series Considered in this Work**

|  | BDA Series | OPV3-2BT-X Series | P5-cyclophane-2SH-X Series |
|---|---|---|---|
| Conductance Ratio | 1.15[b] | 1.43[c] | 5.21[d] |
| $IE_g$ Range[e] (eV) | 0.450 | 0.205 | 0.117 |
| RNE Range[f] (eV) | 0.154 | 0.045 | 0.031 |
| Backbone Length[e] (nm) | 0.57 | 1.96 | 2.37 |

[a]To be comparable, these are for two substituents (-OCH$_3$ and –Cl) that are conserved across all three series
[b]From Venkatarman et al.[17]
[c]From MCBJ experiments in this work
[d]From Lo et al.[16]
[e]From DFT calculations in this work
[f]From image charge model calculations in this work

5. METHODS

**Molecular synthesis.** The seven differently substituted (*E,E*)-OPV3-2BT-X derivatives shown in Figure 1b were synthesized using modified literature procedures (SI section S1.1). The compounds are stable in ambient conditions and have been characterized by NMR spectroscopy (SI section S1.2) and mass spectrometry (SI section S1.3). Each molecule was dissolved in HPLC grade (>99.8%) dichloromethane (DCM) to produce an ~1 μM molecular solution used for deposition on MCBJ samples.

**Single molecule conductance measurements.** The procedure for measuring the single-molecule conductance of each member of the OPV3-2BT-X series is described thoroughly in a previous publication.[27] Briefly, each MCBJ sample was fabricated on a phosphor bronze substrate coated with an insulating layer of polyimide. Electron beam lithography was used to define the pattern of a thin wire with a narrow constriction (~100 nm) in its center. This pattern was then coated with 4 nm of titanium and 80 nm of gold using electron beam metal evaporation, after which reactive ion etching with an $O_2/CHF_3$ plasma was used to create an ~1 μm under-etch beneath the gold bridge in the middle.

MCBJ samples were bent in a custom apparatus employing a push rod driven by both a stepper motor and a piezo actuator. Conductance across the gold bridge was measured by a custom high-bandwidth amplifier based on a Wheatstone bridge.[46] For each sample, an automated LabView program was used to collect thousands of individual breaking traces under ambient conditions. Molecular solutions were drop cast inside a Kalrez gasket in the center of each junction using a clean glass syringe, and multiple molecular datasets were collected from multiple MCBJ samples for each molecule (see Bamberger et al.[27] for our definition of a separate molecular dataset).

Each molecular dataset was independently analyzed using a Segment Clustering algorithm described previously to extract peak single-molecule conductance values.[27] After unambiguously selecting a "main plateau cluster" from each clustering output, the peak conductance for the dataset was determined by fitting an unrestricted Gaussian to the conductance distribution for that cluster on a logarithmic scale. No systematic relationship was observed between peak conductance and molecular concentration, supporting the view that the main plateau cluster corresponds to measuring single- rather than multi-molecule junctions. In our implementation, Segment Clustering produced twelve different peak conductances for each dataset to represent the uncertainty in the cluster bounds. For this work, we took the median value from among the combined set of all twelve peaks from all datasets for each molecule to represent its experimental conductance. The error bars on these conductances represent the range of the middle 68% of peak values in each combined set.

**NEGF-DFT transport calculations.** Each molecular junction was simulated by an OPV3-2BT-X molecule connected to two gold electrodes with finite thickness which taper to single atom apexes. Initial geometry optimizations were performed using density functional theory (DFT) with VASP[47] using a plane wave cutoff of 400 eV and a force convergence threshold of 0.02 eV/Å. Two Au layers at the end of each electrode were restricted from moving during the optimization process to maintain the structural integrity of the electrodes. A region of vacuum was included so that one electrode could be moved relative to the other while maintaining at least 20 Å of separation between periodic images to avoid interactions. Geometry optimization was repeated for a range of electrode separations, and showed that the most stable configuration occurred for an electrode gap size of ~2.3 nm for each molecule (SI section S2).

Optimized geometries from VASP were used in Nanodcal for transport calculations. The junction is seamlessly connected to semi-infinite bulk Au electrodes at zero potential difference. All valence electrons are treated with double-zeta polarization (DZP) basis functions.

To calculate relative conductance for Figure 4, the transmission for each molecule was first divided by the transmission for OPV3-2BT-H at each energy in the shaded region of Figure 3b (-1.5 to -1.0 eV). The median from amongst these values was then used as the relative conductance value for each molecule, with the error bars representing the central 68% of the distribution of ratios for each molecule.

**Local potential calculations.** The $xy$-averaged local electrostatic potential relative to $E_F$ was calculated in VASP for each molecular configuration used in the transport calculations, as well as for an empty junction of the same size. The difference between the empty junction and the junction containing each molecule was then averaged across the length of the molecule to calculate the shift in local potential ($ΔLP$) for each molecule. In order to translate $ΔLP$ into the shift in the local vacuum level

($\Delta V$) used as a reference point for $IE_g$, the shift in electrostatic potential above an isolated molecule ($\Delta MP$) must be excluded, since it is already included in $IE_g$. $\Delta MP$ was estimated using the $xy$-averaged electrostatic potential calculated for each isolated molecule, thus ignoring any charge redistributions due to interactions with the electrodes. The shift in the local vacuum level within the junction is then calculated as $\Delta V = \Delta LP - \Delta MP$. See SI section S5 for further details.

**Image charge calculations.** To calculate RNE values, we employed a simplified model of image charge effects in which each electrode is treated as a perfectly-conducting half-infinite plane and the molecular charge distribution is approximated as a set of point charges centered on each atom. Despite its simplicity, this type of image charge model has provided useful insight into several single-molecule transport results.[14,34,48–51] It is particularly appropriate for the present case because we are focused on differences between similar molecules, and so systematic errors in the model (e.g., from treating the electrodes as planes instead of sharp tips) are likely to cancel. However, we stress that the results of this model should not be quantitatively compared to other results in this work, but are instead only meaningful in terms of revealing qualitative trends and estimating the order of magnitude of RNE values.

For the OPV3-2BT-X series, atomic positions were fixed at the same values used for the transport calculations (see above; separate calculations with the atomic positions free to move were used to calculate vertical $IE_g$ values). Hirshfeld charges on each atom were then calculated for every molecule with a net charge of both 0 and +1 using Gaussian 16 with B3LYP/6-311++G(d,p), in the absence of electrodes. Leaving off the electrodes for these partial charge calculations is justified because we are focused on trends across a series of similar molecules, and moreover the neutral molecules show very minimal charge redistribution when the electrodes are included (SI section S8). Each molecule was then aligned to place its two sulfur atoms along the $x$-axis and the two half-infinite electrodes were placed in the $yz$-plane with the gap size equal to the sulfur-to-sulfur distance of the molecule plus 0.2 nm. This gap size was chosen to roughly account for both the sulfur-to-gold bond lengths and the distance between the gold nuclei and the image plane, but different gap sizes do not change the trend across molecules (SI section S9). Differences in sulfur-to-sulfur lengths between molecules were extremely slight and do not account for the differences in image charge renormalization energy. For both the neutral and cationic form of each molecule, image charges were added until the cumulative interaction energy converged to within 2 meV, and then RNE was calculated as $\Delta E_{cat} - \Delta E_{neu}$ (Figure S41).

For the other molecular series considered in this work, atomic positions were fixed to values determined from DFT gas-phase equilibrium optimization calculations for each neutral molecule performed using Gaussian 16 with B3LYP/6-311++G(d,p). All subsequent steps were the same as for the OPV3-2BT-X series.

After obtaining atomic partial charges and coordinates from DFT results, all image charge calculations were performed using a custom Matlab code. This code has been made freely available online at github.com/LabMonti/ImageChargeCalculations_ForSingleMoleculeJunctions.

## ASSOCIATED CONTENT

**Supporting Information**

The Supporting Information is available free of charge at [doi].

> Molecular synthesis, variation of theoretical results with gap size, conformations of OPV3-2BT-X, impact of twist angles on OPV3-2BT-X conductance, interpreting and referencing vacuum level shifts, justification for simplified conceptual picture of image charge interactions, proof of symmetry of image charge interactions, charge reorganization when molecules are attached to electrodes, gap size for image charge calculations, impact of electron-donating substituents on RNE, image charge correlation in the case of LUMO-mediated transport, further details on molecular series comparisons, and trend in RNE variation with backbone length.

## AUTHOR INFORMATION

Corresponding Authors

>Oliver L.A. Monti: Department of Chemistry and Biochemistry, University of Arizona, Tucson, Arizona 85721, United States; Department of Physics, University of Arizona, Tucson, Arizona 85721, United States; Email: monti@u.arizona.edu; Phone: ++ 520 626 1177; orcid.org/0000-0002-0974-7253.

>Dominic V. McGrath: Department of Chemistry and Biochemistry, University of Arizona, Tucson, Arizona 85721, United States; Email: mcgrath@arizona.edu;

Phone: ++ 520 626 4690; orcid.org/0000-0001-9605-2224.

Manuel Smeu: Department of Physics, Binghamton University–SUNY, Binghamton, NY 13902, USA; Email: msmeu@binghamton; orcid.org/0000-0001-9548-4623.

**Notes**

The authors declare no competing financial interest.

**Author Contributions**

#These authors contributed equally to this work.

ACKNOWLEDGMENTS

The authors would like to acknowledge support from the National Science Foundation award no. DMR-1708443, as well as from the Graduate and Professional Student Council at The University of Arizona. Plasma etching was performed in part using a Plasmatherm reactive ion etcher acquired through an NSF MRI grant, award no. ECCS-1725571, as well as an AGS reactive ion etcher located in the Micro/Nano Fabrication Center at the University of Arizona. Clustering was performed using High Performance Computing (HPC) resources supported by the University of Arizona TRIF, UITS, and RDI and maintained by the UA Research Technologies department. Quality control was performed using a scanning electron microscope (SEM) in the W.M. Keck Center for Nano-Scale Imaging in the Department of Chemistry and Biochemistry at the University of Arizona with funding from the W.M. Keck Foundation Grant. All NMR data were collected in the NMR facility of the Department of Chemistry and Biochemistry at the University of Arizona. The purchase of the Bruker AVANCE III 400MHz spectrometer was supported by the National Science Foundation under Grant Number 840336 and the University of Arizona. NEGF-DFT computations were performed on the Binghamton University HPC cluster, "Spiedie".